# MOBILE-BASED VIDEO CACHING ARCHITECTURE BASED ON BILLBOARD MANAGER

Rajesh Bose[1], Sandip Roy[2] and Debabrata Sarddar[3]

[1, 2, 3] Department of Computer Science & Engineering, University of Kalyani, Kalyani, West Bengal, India
[1] bose.raj00028@gmail.com, [2] sandiproy86@gmail.com, [3] dsarddar1@gmail.com

## ABSTRACT

*Video streaming services are very popular today. Increasingly, users can now access multimedia applications and video playback wirelessly on their mobile devices. However, a significant challenge remains in ensuring smooth and uninterrupted transmission of almost any size of video file over a 3G network, and as quickly as possible in order to optimize bandwidth consumption. In this paper, we propose to position our Billboard Manager to provide an optimal transmission rate to enable smooth video playback to a mobile device user connected to a 3G network. Our work focuses on serving user requests by mobile operators from cached resource managed by Billboard Manager, and transmitting the video files from this pool. The aim is to reduce the load placed on bandwidth resources of a mobile operator by routing away as much user requests away from the internet for having to search a video and, subsequently, if located, have it transferred back to the user.*

## KEYWORDS

*Video-on-Demand, Video Streaming, Multimedia Cache, Data Center, 3G Network*

## 1. INTRODUCTION

Video streaming services are extremely popular today. Today, these services are no longer confined to wired network but are also delivered wirelessly. A growing number of content providers, e.g., YouTube, Netflix, etc. are targeting mobile device users whose numbers are only expected to grow by leaps and bounds. While caching is commonly used by wireless service providers to improve the streaming quality, our proposed Billboard Manager model has been designed keeping in mind cellular and mobile networks with the aim of delivering video content more efficiently over available bandwidth. The proposed Billboard Manager works on focusing on techniques of converting videos to appropriate codecs and resolutions fit for a given mobile device from which request for the video originates. With the Billboard Manager installed, the requested video would be located from a cloud-based node nearest to the location of the user and then stored in its cache. Depending on the model of the mobile device, the Billboard Manager would then transcode the video and begins transmitting it to the mobile device taking into consideration the current connectivity quality [1]. Caching service has been around for some time now. Various techniques exist of cache implementation. By itself, caching is not a complex concept to understand. A cache stores information and data which is frequently accessed and provides fast transfer rates by focusing on optimizing transmission rates. Caches are extensively used on every conceivable electronic gadget or device today. The importance of caching cannot be over-emphasized going by the several hundred million personal computing devices which use processor-based caches [2].

**1.1. How caching service of Billboard Manager works**

We propose to position The Billboard Manager between mobile device clients and cloud video servers. As soon as a request for video is placed to the mobile content service provider, the Billboard Manager scans its cache for the video requested. If the video exists in its cache, the Billboard Manager begins transcoding of the video in the appropriate format suitable for playback on the mobile device requesting the video. Following completion of the process, the Billboard Manager would then begin transferring the video to the mobile device. In case the video does not exist in its cache, the Billboard Manager attempts to locate it from its nearest cloud servers hosting videos. In case the video is found, the Billboard Manager caches it in dedicated cache repository before transcoding and transmitting the video to the mobile device user. As a result of delivering cached content, the mobile content service provider can reduce the quantum of network bandwidth consumption on the server-side, while enhancing end-user experience by offering superior response times to users' requests. Our proposed Billboard Manager has been designed to deliver the following benefits through video caching in the following manner [3]:

**1.2. Quality of experience is optimized:**

Subscribers are able to enjoy the best possible quality of video that are processed for prevailing mobile network conditions.

**1.3. Bandwidth costs are lowered:**

Caching of videos that are looked up by users more than once, translates to reduction in the number of times the videos are looked up and transferred from the respective sites providing such content. This, in turn, decreases the volume of data traversing the networks.

**1.4. Enhanced network efficiency:**

The effect of caching videos significantly improves network conditions by lowering the volume of over-the-top videos transmitted even during peak hours. Consequently, network traffic and/or services that are non-cacheable also see an upward trend in performance.

**1.5. Start up to the playback times is faster:**

Video start times are improved as the content is moved closer to the subscriber. Wait times for the content to be transmitted from distant nodes are also cut down.

**1.6. Video stall reduction:**

Delays arising out of transmission from content servers have little impact on user experience. Video caching services ensure that requested videos are transferred to the respective users over the shortest distance.

**1.7. Streaming Video over 3G Mobile Networks through Billboard Manager**

Our proposed Billboard Manager would employ reduction techniques of reducing resolution size from CIF to QCIF and using fewer frames per second. However, the efficiency of our proposed model would rest not on video compression techniques alone, but on dynamically sending blocks of video in contiguous blocks. The method has been demonstrated by [4].

## 2. RELATED WORK

Video caching is a new research area that has not been sufficiently explored. During recent years, a few commercial video caching systems have been developed. Another body of related work is in the area of scalable video-on-demand systems [5-8]. The idea is to reduce server load by grouping multiple requests within a time-interval and serving the entire group in single stream .The Middle Man architecture is a collection of cooperative proxy servers that collectively act as a video cache for a well-provisioned local area network [9]. Video streams are stored across multiple proxies where they can be replaced at a granularity of a block .They examine performance of the Middle Man architecture with different replacement policies. Other related work has been done on memory caching for multimedia servers [10, 11]. While the basic principle of caching data in different memory levels of a video server has some similarities with storing data in a distributed caching system, there is a fundamental difference. The spatial distance between different memory levels in a server is zero. In contrast, spatial distance between distributed caching systems is not negligible and, therefore, has to be considered in the design of web cache management policies. A good report has been presented in [12]. Another good report on mobile cloud architecture that helps us to solve the caching problem is presented in [13]. Ref. [14] a proxy caching mechanism is used for improving delivered quality of layered encoded multimedia streams.

## 3. PROPOSED WORK

A mobile user sends request for video to the Billboard Manager. The request is checked against its index table. If the request for content is not found in its index table, the Billboard Manager passes an HTTP byte range request to retrieve the content from one of the registered cloud nodes of the Billboard Manager. While retrieval, a copy of this video is saved in its own database to serve identical requests in future. Alternatively, if the request for video can be served directly, the Billboard Manager is able to deliver the video direct from its own cached resource thereby reducing the processing time which would have been otherwise expended in searching. The Billboard Manager also uses a video converter which is a part of its own system. The transcoder transforms video from online video streaming sites, e.g., YouTube, Daily motion, etc., into the requisite format playable on mobile devices. The Billboard Manager also splits the encoded streams into segments and sends each converted stream segments to mobile devices over mobile or local networks. The Billboard Manager acts as an interface between users of mobile devices and the cloud. Our proposed model involving Billboard Manager has been designed keeping in mind saving of 3G network bandwidth at either the user and/or the cloud service provider ends. The sole purpose of our proposed model is to achieve a considerable reduction in data transmission across any given 3G mobile networks from cloud-based nodes. The end goal is to achieve direct savings in terms of costs and bandwidth resources utilized. Our model suggests an architecture, wherein, the Billboard Manager is able to provide video streams directly to a 3G mobile device user from its own cached resource rather than accessing the same video from across one of its own registered cloud nodes.

This directly translates in time savings as well. In the absence of a video not in its cached resource, the 3G mobile user requesting the service would be served with a controlled video stream. The mechanism of this would be managed in a manner such that the Billboard Manager initially begins caching chunks of the video from its own cloud-based nodes, and then sends from it, smaller streams to the 3G mobile device user. In this manner, the 3G mobile device user, after experiencing an initial time delay from the moment the user places the request to the time when the video starts on the device, enjoys an uninterrupted video stream thereon. Further, once a video is fully cached, the Billboard Manager can serve it from its resource to other mobile device users placing a request for it. This reduces overall bandwidth consumption as the Billboard Manager does not have to engage resources to obtain the video once again from its registered cloud-based nodes. The 3G mobile device user requesting the video only has to use their own (3G) network bandwidth and can do so with maximum effect with little or no delay between placing the request to the time the video starts playing. Thus, with the aid of this proposed model, a considerable amount of resources in terms of bandwidth, cost and time can be saved. Figure 1 shows the function of our proposed model.

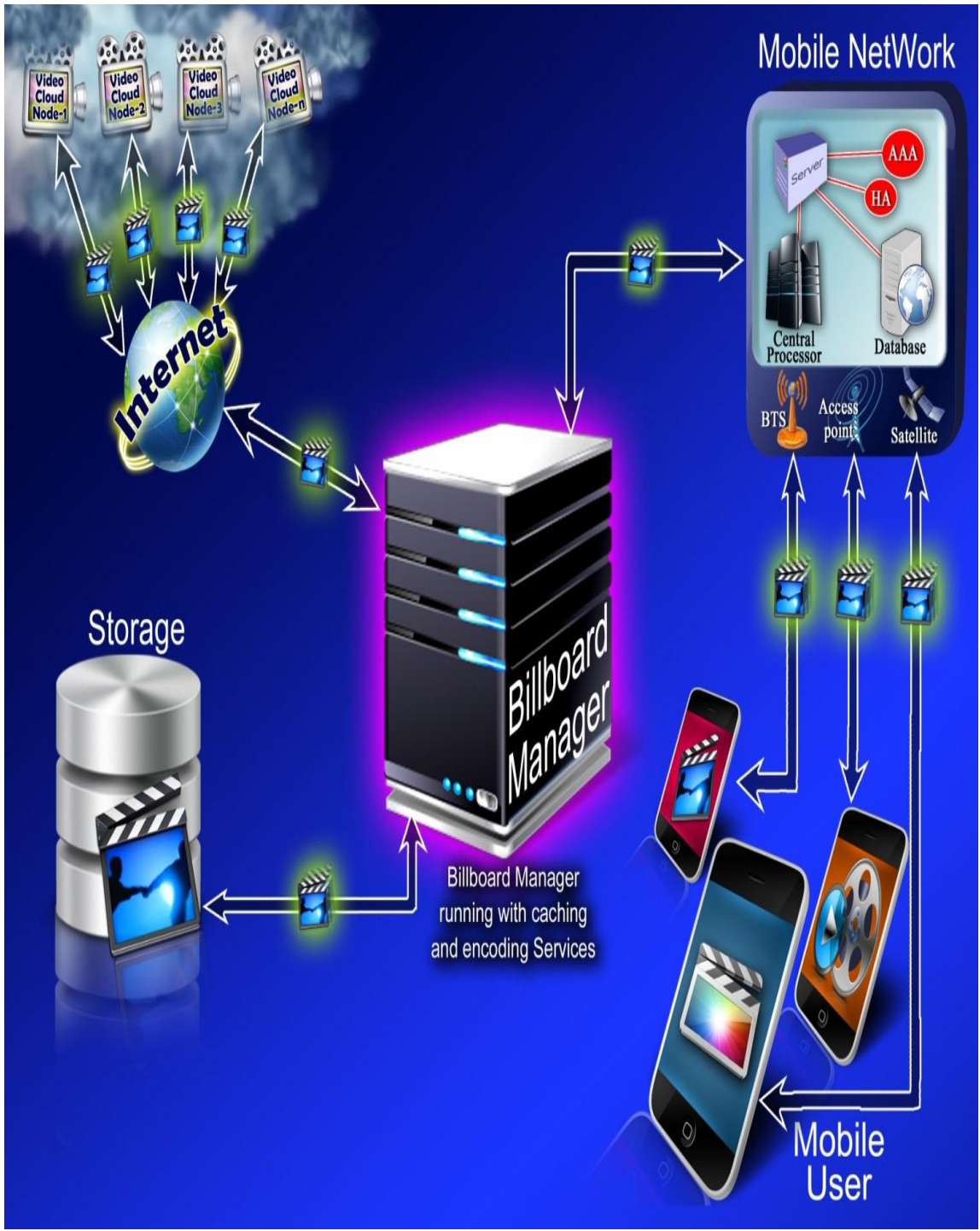

Figure 1.  Proposed Caching Architecture

### 3.1. Algorithm of our proposed cloud architecture

1. User places a request to play a video.
2. The Billboard Manager takes in the request. All such requests, in our proposed model, is routed through the Billboard Manager.

3. The request is processed against an index maintained by the Billboard Manager. The index contains details of videos that can be served by the Billboard Manager from its own cached resources and those from its registered cloud-based nodes.
4. If the requested video is found in its own cached resources, the Billboard Manager begins streaming it to the user requesting the video after processing it as described in the following steps #5 through 8. If not, the Billboard Manager moves forward to step #11 and #12.
5. The Billboard Manager checks whether the video is in the appropriate format required for playing on the mobile device of the user requesting the video.
6. If it is, it proceeds to step #8. If not, it proceeds with step #7, i.e., the following step.
7. The Billboard Manager converts the video to an appropriate format that can be streamed directly to the mobile device for instant playback.
8. The Billboard Manager splits the encoded video into segments. Each such segment is then transmitted to the mobile device over 3G mobile network.
9. After complete transfer of the file, the Billboard Manager stands by for the next request from a mobile device user.
10. If the requested video is not listed in its index, the Billboard Manager begins to lookup another set of index which details availability of the video at the registered cloud-based nodes of the Billboard Manager.
11. If the requested video cannot be found either in its cached resources or with any of its registered cloud-based nodes, the Billboard Manager notifies the user, skips all the following steps and awaits further request from the mobile device user. Otherwise, it proceeds with the next step of retrieving the video file from its registered cloud-based nodes.
12. If the Billboard Manager finds the video to be stored at more than one cloud-based node, it begins to select the best among them from where the video file can be transmitted to the Billboard Manager. The process of selection is based on several factors which are outlined in the next steps.
13. All the cloud-nodes registered with the Billboard Manager keep sending information at periodic intervals. The information consists of network channel capacity and available storage space.
14. The Billboard Manager constantly computes an index score for each of the registered cloud-based nodes. The index score is computed taking into account the constant periodic inputs from the nodes themselves, and the shortest route to each of the nodes.
15. At a given time when the Billboard Manager is about to select the best fit cloud-based node from where the video file is to be streamed back to it, the following steps are followed which are inherent to the steps #13 and 14 above. In other words, an instant index score is calculated based on the following parameters and comparisons.
    a. Select the shortest route to the cloud-based nodes hosting the video file.
    b. If more than one node shares identical route times, the channel capacity is compared. Otherwise, transmission of the video file begins from the cloud-based node to the Billboard Manager.
    c. After channel capacities are compared, the one with the maximum value is selected for the video file to be transmitted. Otherwise signal strengths are compared.
    d. The node with the best signal strength is selected and the video transmission takes place between the corresponding cloud-based node and the Billboard Manager.
    e. In case, after going through all the comparisons, more than one cloud-based nodes are identically matched to transmit the video file, one of them is chosen at random by the Billboard Manager.
16. Upon completion of transfer of the video, the Billboard Manager proceeds with step #5 onward.
17. The algorithm ends with the Billboard Manager completing the workflow cycle at steps #9. In the event, the video file does not exist within its registered network of cloud-based nodes, or within its own cached resources, step #11 is executed.

## 3.2. Flowchart of our proposed cloud architecture

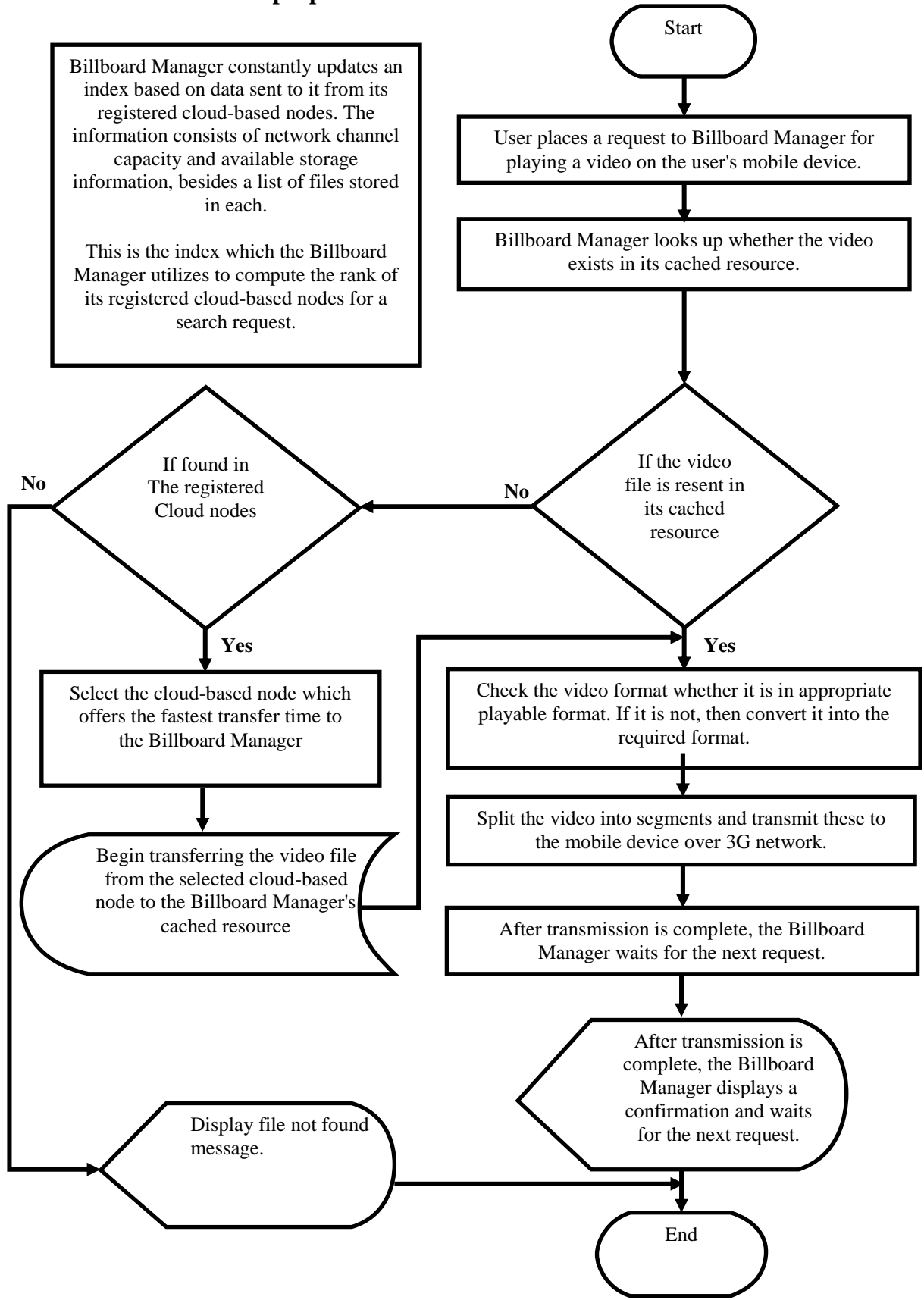

## 4. CONCLUSIONS

This paper describes a proposed design of mobile video architecture based on our projected Billboard Manager. The algorithm detailed in this paper explains the steps that the Billboard Manager intends to take to create an efficient environment within which both the service provider and the user are able to optimize available bandwidth resources to the extent possible. Our proposed model would help send video streams at appropriate frame rates in a relatively short span of time, suitable for the mobile device requesting the video file. This is achieved with the help of caching and encoding services. In the initial stage, when the requested video is not available in the local cache, there would be a certain amount of time which would be involved to fetch it from the appropriate cloud node registered with our proposed Billboard Manager. Once the video is fetched, a copy of it would be stored automatically in its local cache by the Billboard Manager to serve subsequent requests. As a result, many resources in the form of bandwidth and time can be saved in cases where requests for the identical video file are ever placed in future. This proposed model wherein the Billboard Manager utilizes local cache database instead of repeatedly engaging bandwidth necessary to transfer files from cloud nodes to mobile device users, would help efficiently reduce video playback lag from the time it takes for the user to complete placing the request for the video.

## ACKNOWLEDGEMENTS

The authors express their gratitude towards staff and members of the Department of Computer Science & Engineering, University of Kalyani for helping in arranging computation resources that have been used in the work.

**Authors**

Rajesh Bose is currently pursuing Ph.D from University of Kalyani. He is an IT professional employed as Senior Project Engineer with Simplex Infrastructures Limited, Data Center, Kolkata. He received his degree in M.Tech. in Mobile Communication and Networking from WBUT in 2007. He received his degree in B.E. in Computer Science and Engineering from BPUT in 2004. He has also several global certifications under his belt. These are CCNA, CCNP-BCRAN, and CCA (Citrix Certified Administrator for Citrix Access Gateway 9 Enterprise Edition), CCA (Citrix Certified Administrator for Citrix Xen App 5 for Windows Server 2008). His research interests include cloud computing, wireless communication and networking.

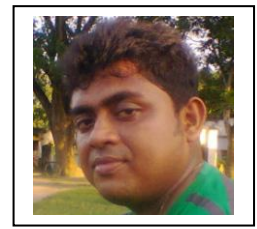

Sandip Roy is currently pursuing Ph.D from University of Kalyani. He is an Assistant Professor in the Department of Information Technology, Brainware Group of Institutions, Kolkata, West Bengal, India. He has completed M.Tech in Computer Science & Engineering from HIT under WBUT in 2011. He has also done his B.Tech in Information Technology from WBUT in 2008. His main areas of research interest are Cloud Computing, Data Structure and Algorithm.

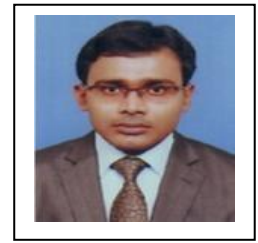

Debabrata Sarddar is an Assistant Professor in the Department of Computer Science and Engineering, University of Kalyani, Kalyani, Nadia, West Bengal, INDIA. He has done PhD at Jadavpur University. He completed his M. Tech in Computer Science & Engineering from DAVV, Indore in 2006, and his B.E in Computer Science & Engineering from NIT, Durgapur in 2001. He has published more than 75 research papers in different journals and conferences. His research interest includes wireless and mobile system and WSN, Cloud computing.

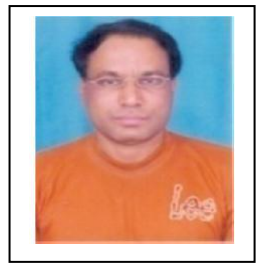